\documentclass[a4paper,11pt]{article}
\usepackage{pos}
\usepackage{xspace}

\newcommand{\TeV}{{\ensuremath\rm TeV}\xspace}
\newcommand{\GeV}{{\ensuremath\rm GeV}\xspace}
\newcommand{\fb}{{\ensuremath\rm fb}\xspace}

\newcommand{\eqn}{equation}
\newcommand{\lam}{\lambda}

\newcommand{\lb}{\left(}
\newcommand{\rb}{\right)}

\title{Models with (broken) $\mathbb{Z}_2$ symmetries}

\author*[a,b]{Tania Robens}

\affiliation[a]{Division of Theoretical Physics, Rudjer Boskovic Institute\\
  Bijenicka cesta 54, 10000 Zagreb, Croatia}

\affiliation[b]{Theoretical Physics Department, CERN\\ 1211 Geneva 23, Switzerland}

\emailAdd{trobens@irb.hr}

\abstract{I discuss the allowed parameter space as well as collider prospects of several new physics models containing (broken) $\mathbb{Z}_2$ symmetries. I focus on effects of current constraints on these, and touch on several future collider options.\\ RBI-ThPhys-2022-05, CERN-TH-2022-019}

\FullConference{%
  7th Symposium on Prospects in the Physics of Discrete Symmetries (DISCRETE 2020-2021)\\
  29th November - 3rd December 2021\\
 Bergen, Norway}

\begin{document}
\maketitle

\section{Introduction}
In this manuscript, I present several models containing (unbroken) $\mathbb{Z}_2$ symmetries. I focus on the effects of current constraints on these models as well as predictions for current and future collider machines. More detailed discussions can be found e.g. in \cite{Ilnicka:2015jba,Ilnicka:2018def,Kalinowski:2018ylg,Kalinowski:2020rmb} for the Inert Doublet Model (IDM) and \cite{Robens:2019kga,Papaefstathiou:2020lyp} for the Two Real Singlet Model (TRSM), respectively.

\section{Inert Doublet Model}
\begin{figure}
\begin{minipage}{0.49\textwidth}
\begin{center}
\includegraphics[width=0.9\textwidth]{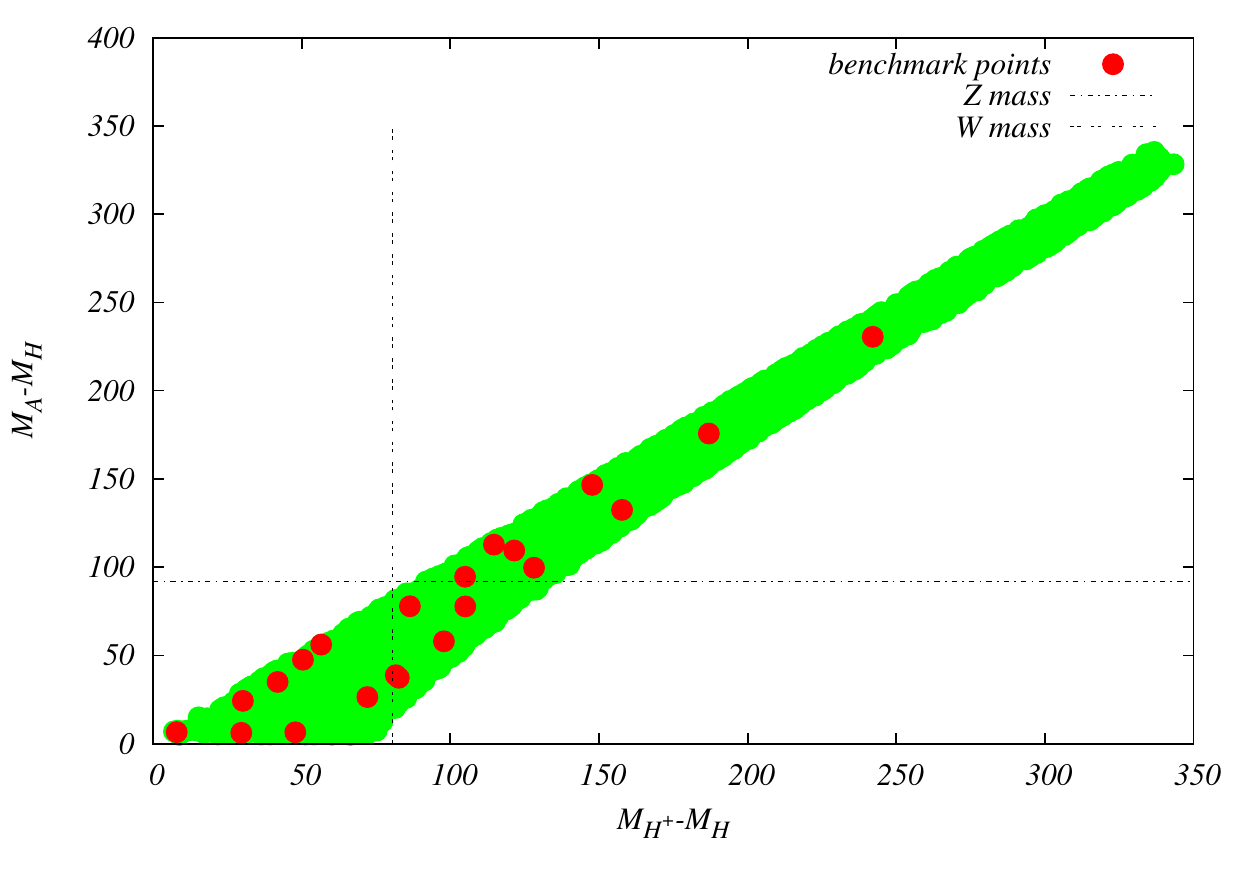}
\end{center}
\end{minipage}
\begin{minipage}{0.49\textwidth}
\begin{center}
\includegraphics[width=0.9\textwidth]{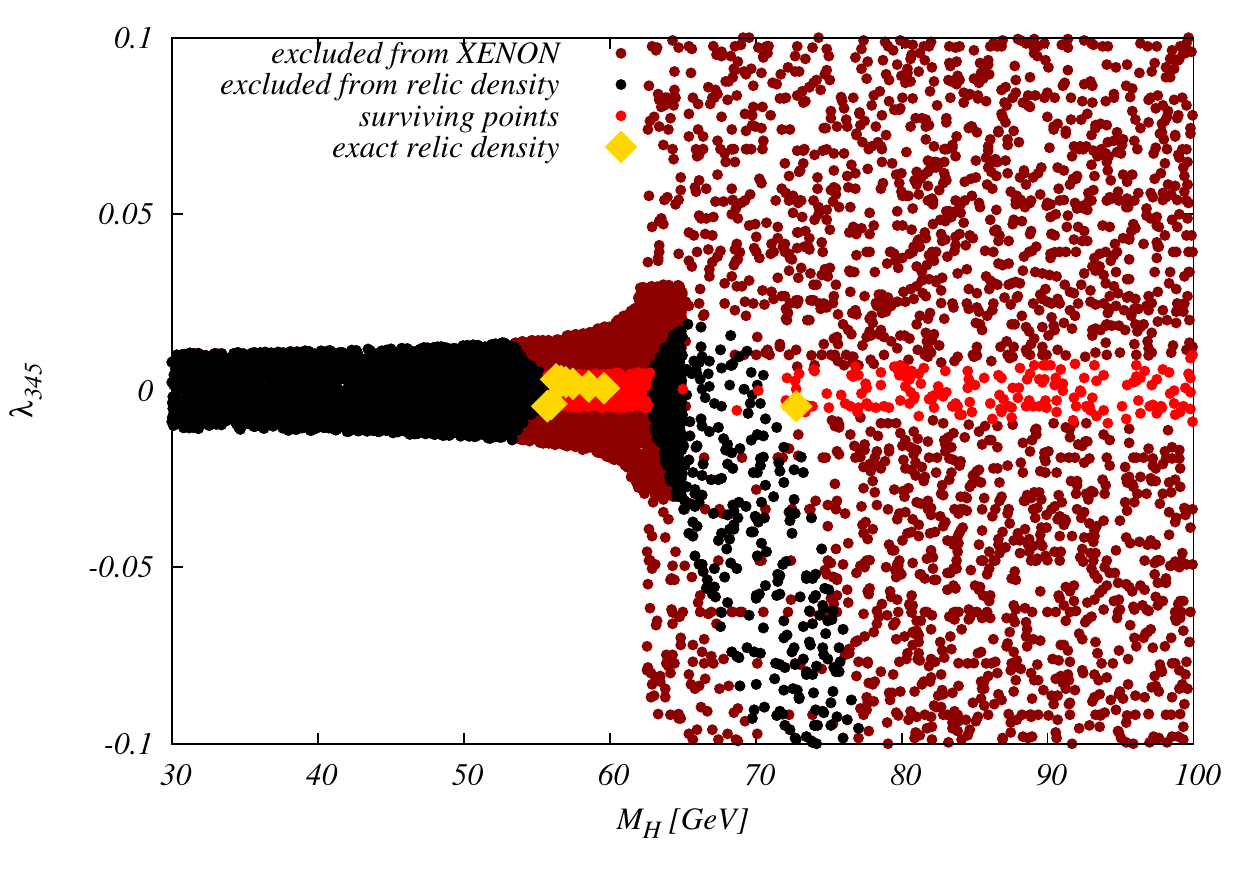}
\end{center}
\end{minipage}
\caption{\label{fig:idmscan} {\sl Left:} Masses are requested to be quite degenerate after all constraints have been taken into account. In the {$\lb M_{H^\pm}-M_H,\,M_A-M_H \rb$} plane (taken from \cite{Kalinowski:2018ylg}). {\sl Right:} Interplay of signal strength and relic density constraints in the $\lb M_H,\,\lam_{345}\rb$ plane, using XENON1T results, with golden points labelling those points that produce exact relic density (taken from \cite{Ilnicka:2018def}).}
\end{figure}

The Inert Doublet Model is a two Higgs doublet model (2HDM) that obeys a discrete $\mathbb{Z}_2$ symmetry that is unbroken, inducing a dark matter candidate that stems from the second doublet \cite{Deshpande:1977rw,Cao:2007rm,Barbieri:2006dq}. The model features four additional scalar states $H,\,A,\,H^\pm$, and has in total 7 free parameters prior to electroweak symmetry breaking
\begin{\eqn}
v,\,m_h,\,\underbrace{m_H,\,m_A,\,m_{H^\pm}}_{\text{second doublet}},\,\lam_2,\,\lam_{345}\,\equiv\,\lam_3+\lam_4+\lam_5,
\end{\eqn}
where the $\lam_i$s are standard couplings appearing in the 2HDM potential. Two parameters ($m_h$ and $v$) are fixed by current measurements. The model is subject to a large number of experimental and theoretical constraints \cite{Ilnicka:2015jba,Ilnicka:2018def,Kalinowski:2018ylg,Kalinowski:2020rmb,Robens:2021yrl}. A general feature is a relatively strong degeneracy between the additional masses of the second doublet, as well as a minimal mass scale for the dark matter candidate resulting from a combination of relic density and signal strength measurement constraints (see \cite{Ilnicka:2015jba,Kalinowski:2020rmb} for a detailed discussion). These features are displayed in figure \ref{fig:idmscan}.

\subsection{Sensitivity study at current and future colliders}

\begin{figure}
\begin{center}
\begin{minipage}{0.42\textwidth}
\begin{center}
\includegraphics[width=\textwidth]{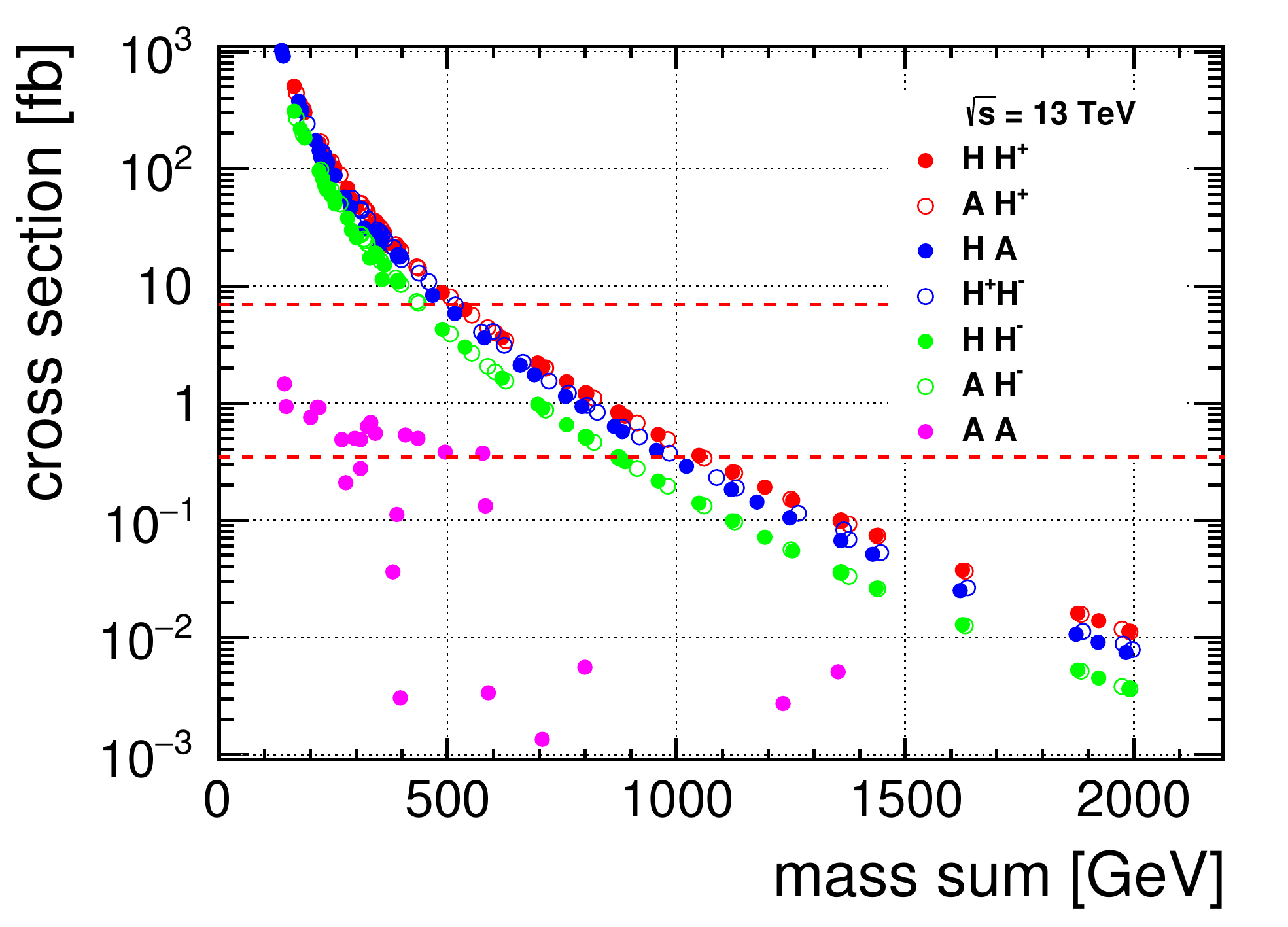}
\end{center}
\end{minipage}
\begin{minipage}{0.42\textwidth}
\begin{center}
\includegraphics[width=\textwidth]{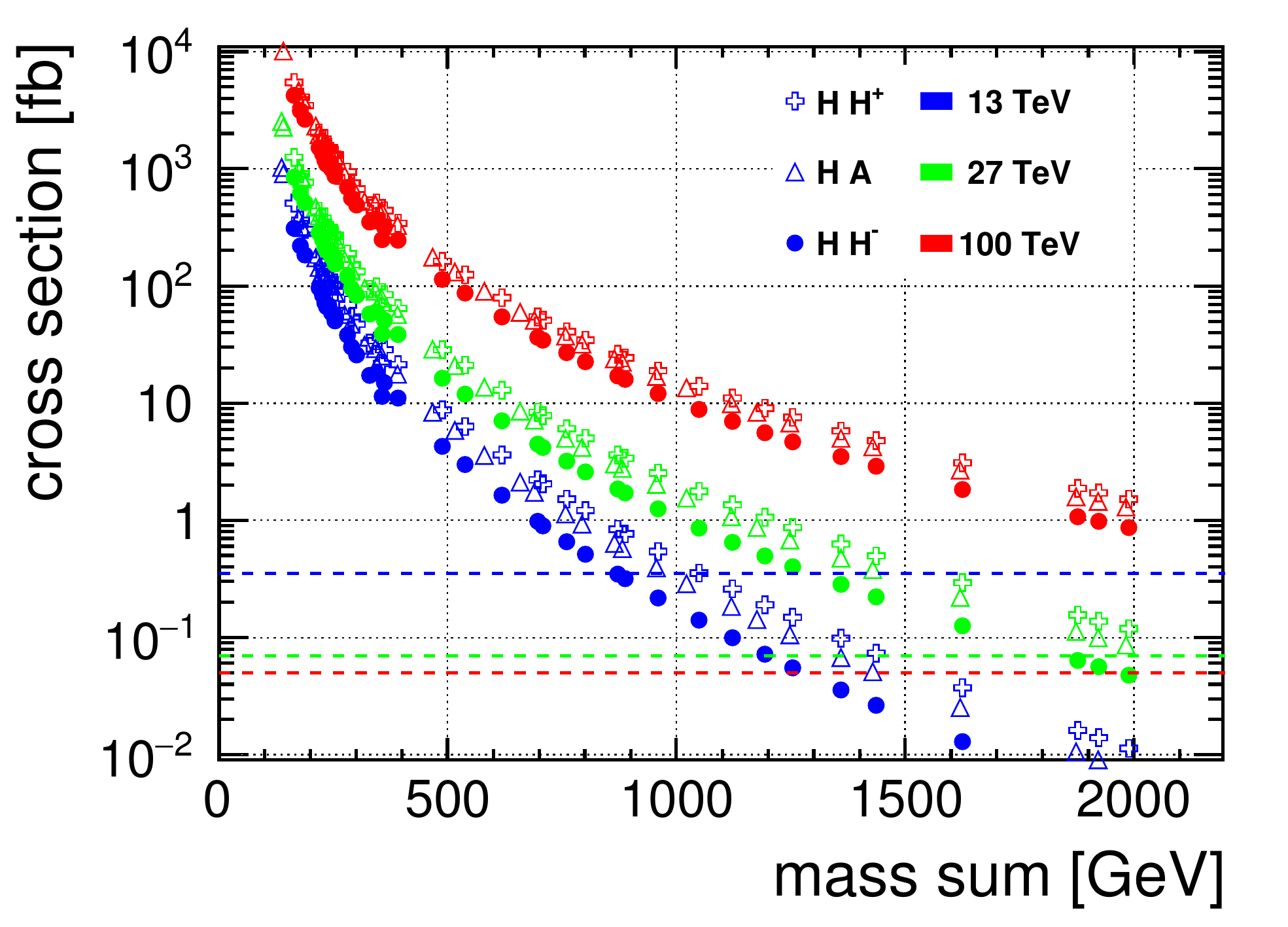}
\end{center}
\end{minipage}
\begin{minipage}{0.42\textwidth}
\begin{center}
\includegraphics[width=\textwidth]{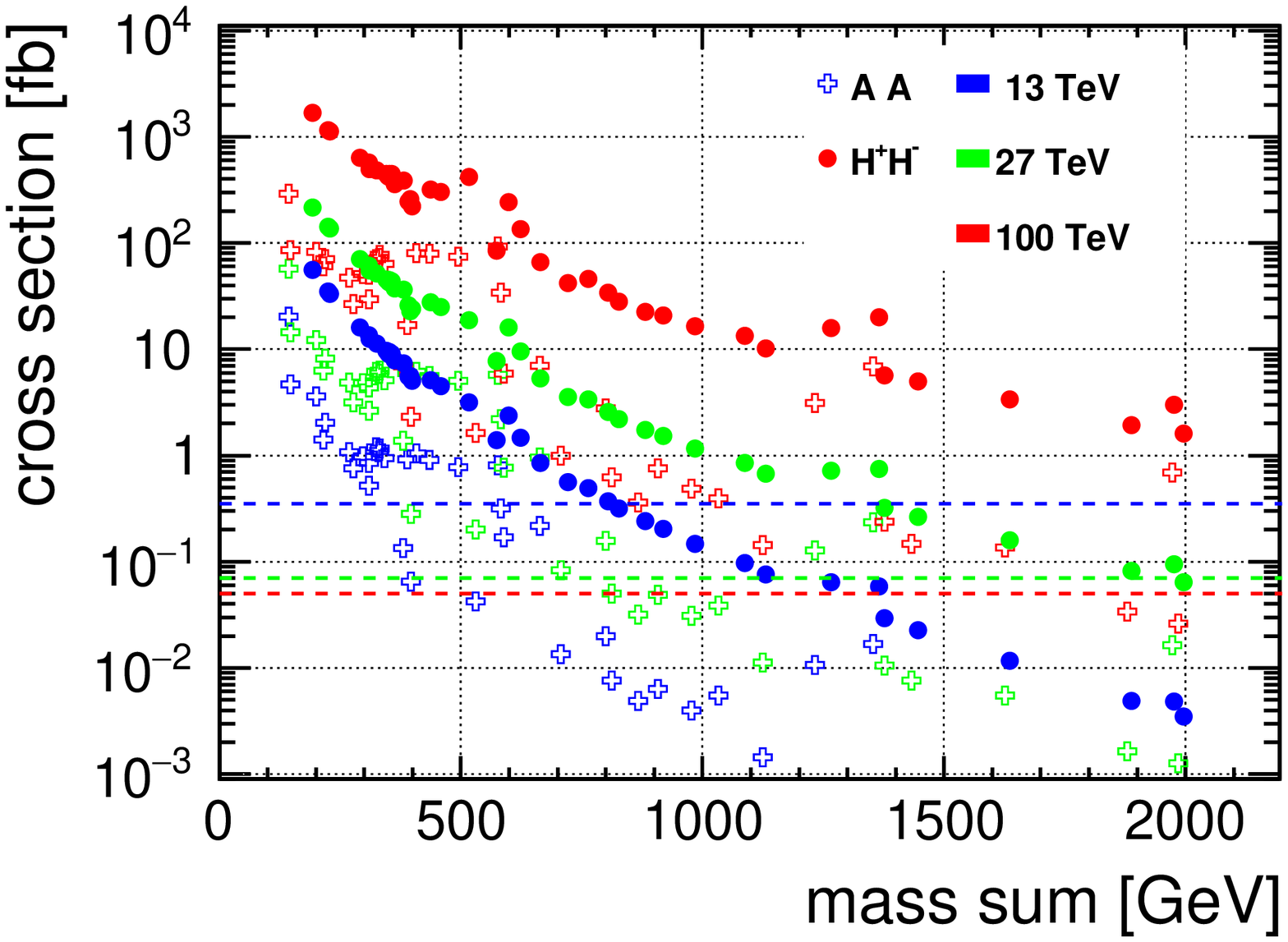}
\end{center}
\end{minipage}
\begin{minipage}{0.42\textwidth}
\begin{center}
\includegraphics[width=\textwidth]{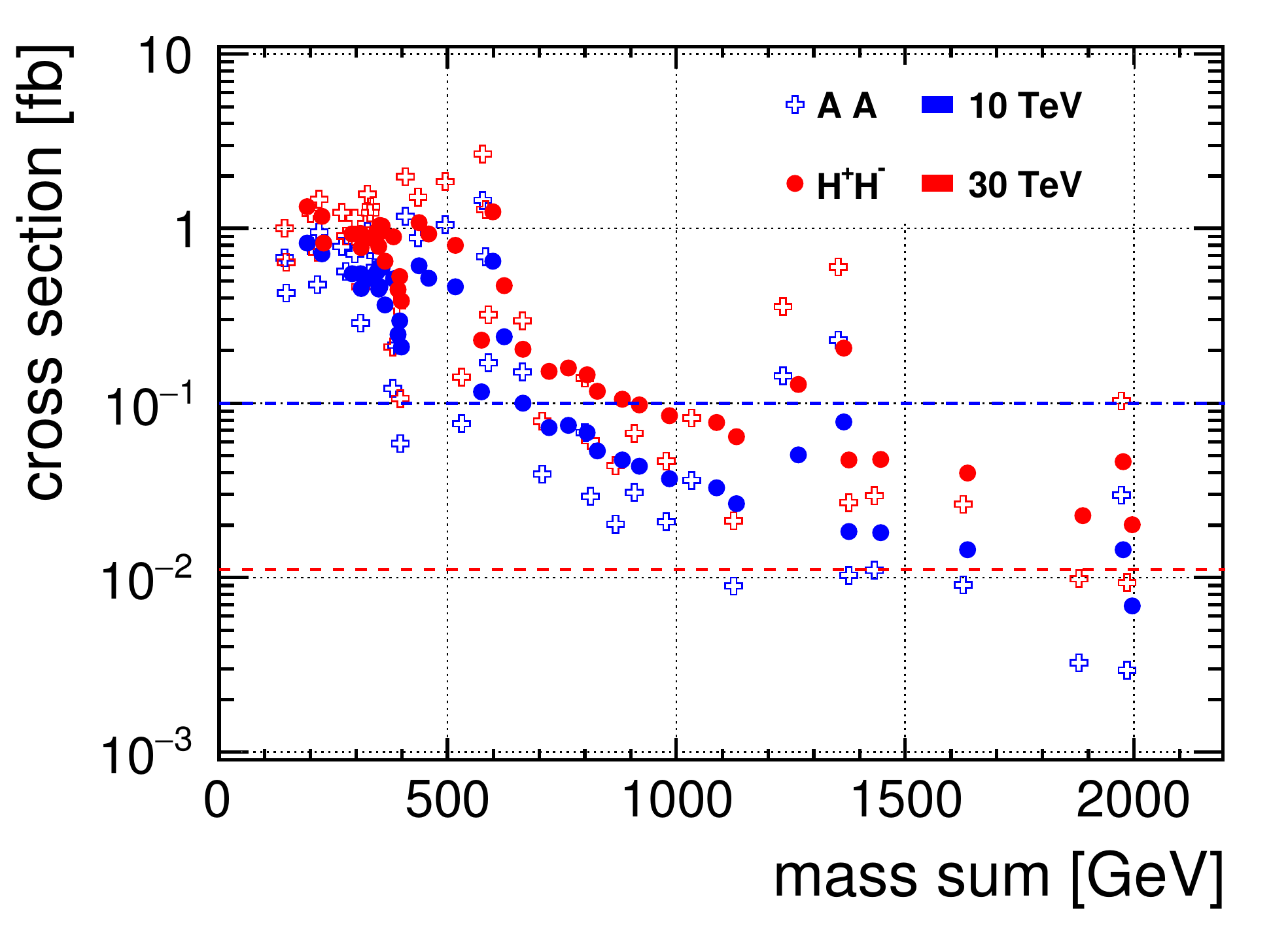}
\end{center}
\end{minipage}
\end{center}
\caption{\label{fig:idm} Predictions for production cross sections for various processes and collider options. {\sl Top left:} Predictions for various pair-production cross sections for a $pp$ collider at 13 \TeV, as a function of the mass sum of the produced particles. {\sl Top right:} Same for various center-of-mass energies. {\sl Bottom left:} VBF-type production of $AA$ and $H^+\,H^-$ at various center-of-mass energies for $pp$ colliders. {\sl Bottom right:} Same for $\mu^+\mu^-$ colliders. Taken from \cite{Kalinowski:2020rmb}. The lines correspond to the cross-sections required to prodce at least 1000 events using the respective design luminosity.}
\end{figure}

I here present the results derived in \cite{Kalinowski:2020rmb}. In that work, a sensitivity comparison for selected benchmark points \cite{Kalinowski:2018ylg,Kalinowski:2018kdn,Kalinowski:2020rmb} using a simple counting criteria was used, where a benchmark point is considered reachable if at least 1000 signal events are produced using nominal luminosity of the respective collider (c.f. also \cite{Robens:2021zvr}). The results are presented in table \ref{tab:sens}, with the accompagnying figures, displaying production cross sections for pair-production of the novel scalars at various collider options and center-of-mass energies in figure \ref{fig:idm}, taken from \cite{Kalinowski:2020rmb}. We here have used Madgraph5 \cite{Alwall:2011uj} with a UFO input file from \cite{Goudelis:2013uca} for cross-section predictions. Results for CLIC were taken from \cite{Kalinowski:2018kdn,deBlas:2018mhx}.

\begin{center}
\begin{table}
\begin{center}
\begin{tabular}{||c||c||c||c||} \hline \hline
{collider}&{all others}& { $AA$} & {$AA$ +VBF}\\ \hline \hline
HL-LHC&1 \TeV&200-600 \GeV& 500-600 \GeV\\
HE-LHC&2 \TeV&400-1400 \GeV&800-1400 \GeV\\
FCC-hh&2 \TeV&600-2000 \GeV&1600-2000 \GeV\\ \hline \hline
CLIC, 3 \TeV&2 \TeV &- &300-600 \GeV\\
$\mu\mu$, 10 \TeV&2 \TeV &-&400-1400 \GeV\\
$\mu\mu$, 30 \TeV&2 \TeV  &-&1800-2000 \GeV \\ \hline \hline
\end{tabular}
\end{center}
\caption{Sensitivity of different collider options, using the sensitivity criterium of 1000 generated events in the specific channel. $x-y$ denotes minimal/ maximal mass scales that are reachable.}
\label{tab:sens}
\end{table}
\end{center}
\section{Two Real Singlet Model}
I now turn to the model introduced in \cite{Robens:2019kga}. In this model, the scalar sector of the Standard Model (SM) is extended by two real scalars obeying a discrete $\mathbb{Z}_2\,\otimes\,\mathbb{Z}_2'$ symmetry. Both fields acquire a vacuum expectation value (vev), softly breaking the above symmetries and leading to mixing between all scalar states. The model is characterized by 9 parameters after electroweak symmetry breaking, $m_1,\,m_2,\,m_3,\,v,\,v_X,\,v_S,\,\theta_{hS},\,\theta_{hX},\,\theta_{SX}$, where $m_i,\,v,\,\theta$ denote masses\footnote{We use the convention $m_1\,\leq\,m_2\,\leq\,m_3$.}, vevs, and mixing angles. One mass $m\,\sim\,125\,\GeV$ and $v\,\sim\,246\,\GeV$ are fixed by current measurements. 

In \cite{Robens:2019kga}, various benchmark planes (BPs) where proposed within this model, allowing for novel production and decay processes, including decay chains which by that time had not been investigated by the LHC experiments, $p\,p\,\rightarrow\,h_3\,\rightarrow\,h_1\,h_2,\;p\,p\,\rightarrow\,h_a\,\rightarrow\,h_b\,h_b$,
where for the symmetric decays we assume none of the scalars corresponds to the SM-like 125 \GeV~ resonance.
\subsection{$h_{125}h_{125}h_{125}$ at the (HL) LHC}
I now focus on one particular benchmark plane (BP3), that features the first production mode, in the scenario with $h_1\,\equiv\,h_{125}$. Depending on $m_2$, this allows for a $h_{125}\,h_{125}\,h_{125}$ final state. For cases where the 125 \GeV scalar exclusively decays into $b\,\bar{b}$ final states, we have conducted a complete phenomenological study for a 14 \TeV~ LHC \cite{Papaefstathiou:2020lyp}.  We made use of a customized  \texttt{loop\_sm} model implemented in \texttt{MadGraph5\_aMC@NLO} (v2.7.3)~\cite{Alwall:2014hca,Hirschi:2015iia}, subsequently interfaced to \texttt{HERWIG} (v7.2.1)~\cite{Bahr:2008pv,Gieseke:2011na,Arnold:2012fq,Bellm:2013hwb,Bellm:2015jjp,Bellm:2017bvx,Bellm:2019zci}. Results are shown in table \ref{tab:hhh}. We see that several benchmark points are already accessible with a relatively low integrated luminosity.
{\small
\begin{table}
\begin{center}
{\small
\begin{tabular}{c||cc||cc}\\
{\bf $(M_2, M_3)$}& $\sigma(pp\rightarrow h_1 h_1 h_1)$ &
$\sigma(pp\rightarrow 3 b \bar{b})$&$\text{sig}|_{300\rm{fb}^{-1}}$& $\text{sig}|_{3000\rm{fb}^{-1}}$\\
${[\GeV]}$ & ${[\fb]}$  & ${[\fb]}$ & &\\
\hline\hline
$(255, 504)$ & $32.40$ & $6.40$&$2.92$&{  $9.23$}\\
$(263, 455)$ & $50.36$ & $9.95$&{  $4.78$}&{  $15.11 $}\\
$(287, 502)$ & $39.61$ & $7.82$&{  $4.01$} &{  $12.68$}\\
$(290, 454)$ & $49.00$ & $9.68$&{  $5.02$}&{  $15.86 $}\\
$(320, 503)$ & $35.88$& $7.09$& {  $3.76 $}&{  $11.88$}\\
$(264, 504)$ & $37.67$ & $7.44$&{  $3.56 $}&{  $11.27 $}\\
$(280, 455)$& $51.00$ & $10.07$&{  $5.18$} &{  $16.39$}\\
$(300, 475)$&$43.92$& $8.68$&{  $4.64 $}&{  $14.68 $}\\
$(310, 500)$& $37.90$ & $7.49$&{  $4.09 $}&{  $12.94$}\\
$(280, 500)$& $40.26$& $7.95$&{  $4.00 $}&{  $12.65 $}\\
\end{tabular}
}
\end{center}
\caption{\label{tab:hhh} 6 b final state {leading-order} production cross sections at 14 \TeV, as well as significances for different integrated luminosities. Taken from \cite{Papaefstathiou:2020lyp}.}
\end{table}
}
\subsection{Recasting current LHC searches}
\begin{figure}
\begin{center}
\begin{minipage}{0.45\textwidth}
\includegraphics[width=\textwidth]{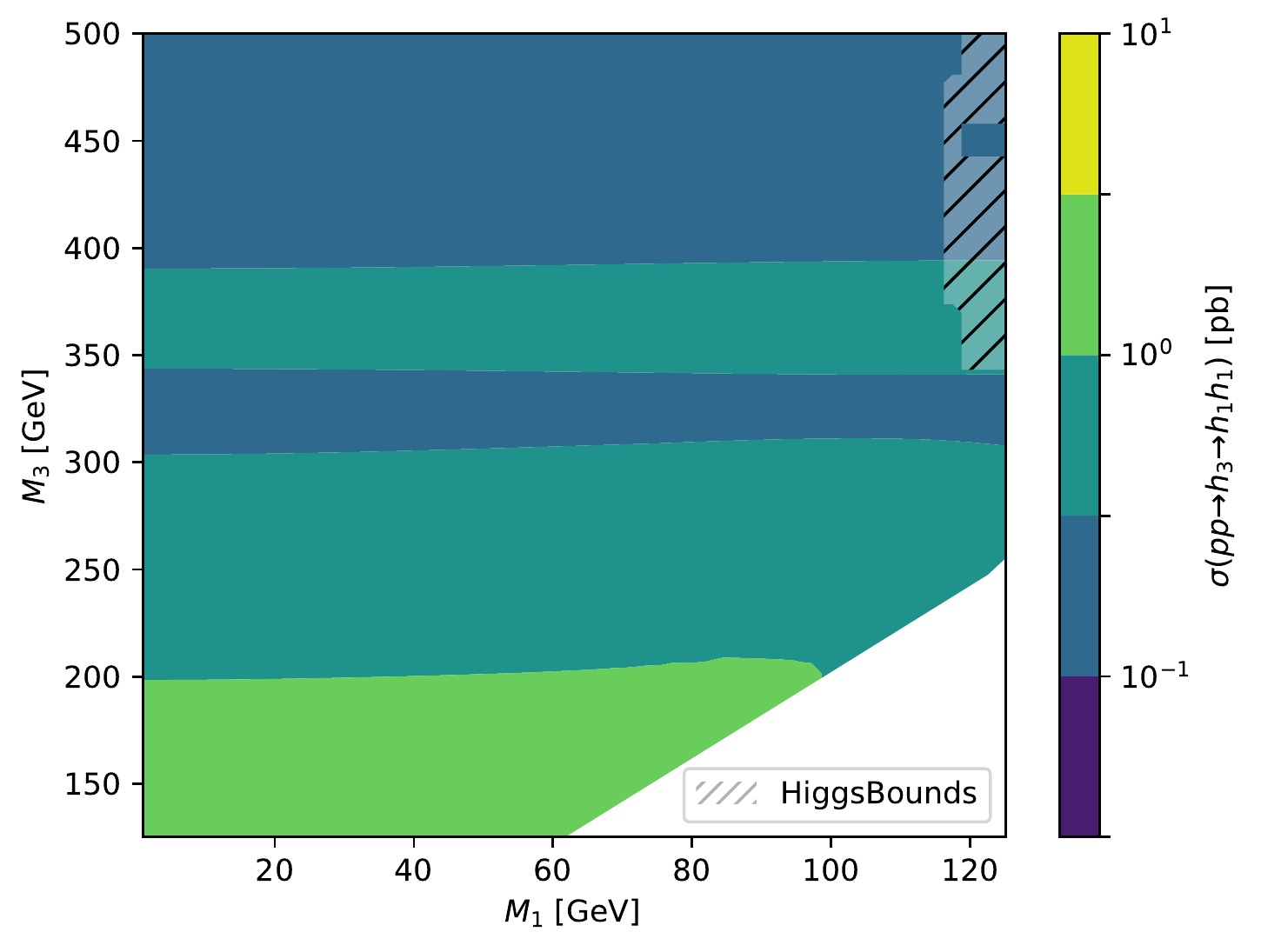}
\end{minipage}
\begin{minipage}{0.45\textwidth}
\includegraphics[width=\textwidth]{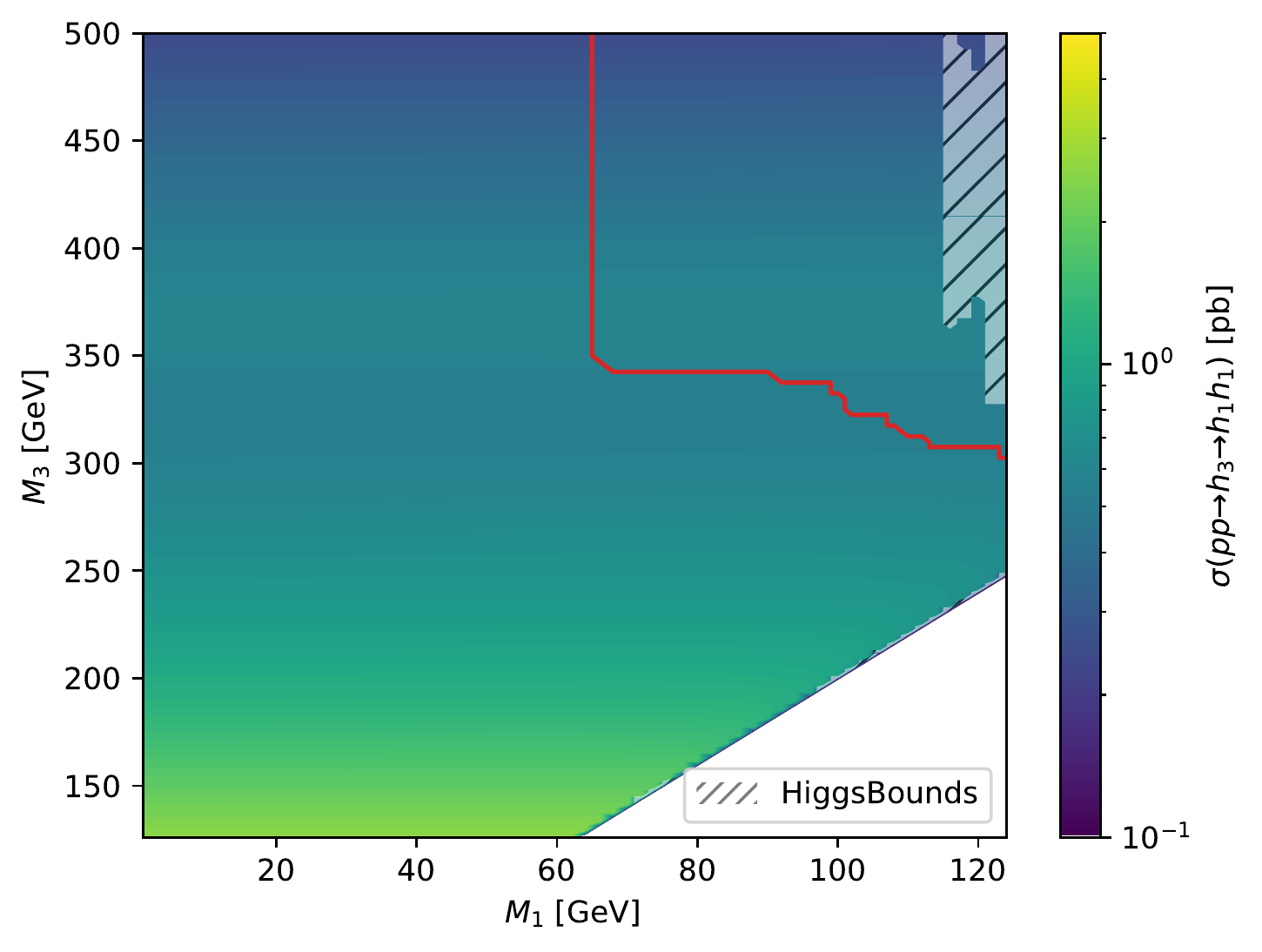}
\end{minipage}
\end{center}
\caption{\label{fig:bp5reint} Reinterpretation of a $36\,\fb^{-1}$ CMS search for di-Higgs production via a heavy resonance using the 4 b final state. The exclusion line uses the results obtained in \cite{Barducci:2019xkq}. Points to the right and above the red contour are excluded.}
\end{figure}

It is also interesting to investigate whether current searches can be reinterpreted and recasted in such a way that they allow to exclude regions in the models parameter space that were not directly scrutinized in the experimental search, or for which no interpretation was presented in the original publication. In \cite{Barducci:2019xkq}, the authors have reinterpreted a CMS search for $p\,p\,\rightarrow\,H\,\rightarrow\,h_{125}h_{125}\,\rightarrow\,4\,b$ \cite{CMS:2018qmt}, which corresponds to di-Higgs production via a heavy resonance and subsequent decays into $b\,\bar{b}$ final states, and extended the mass ranges for the scalars in the decay chain. These results are directly applicable in the TRSM, in particular to BP5 that was designed to focus on $h_3\,\rightarrow\,h_1\,h_1$, where now $h_2\,\equiv\,h_{125}$ represents the 125 \GeV scalar discovered at the LHC. We display the corresponding results in figure \ref{fig:bp5reint}\footnote{We thank the authors of \cite{Barducci:2019xkq} for providing us with the corresponding exclusion limits.}. We see that the sensitive region of parameter space is significantly extended.
\section{Conclusion}
I presented two models with discrete $\mathbb{Z}_2$ symmetries, which in one case were softly broken by the vevs of the respective fields. We have briefly discussed perspectives at (future) colliders for both models, and strongly encourage the experimental collaborations to pursue more detailed studies based on our benchmark scenarions.
\section*{Acknowledgements}
This research was supported in parts by the National Science Centre, Poland, the HARMONIA
project under contract UMO-2015/18/M/ST2/00518 (2016-2021), OPUS project under contract
UMO-2017/25/B/ST2/00496 (2018-2021), COST actions CA16201 - Particleface and  CA15108 - Fundamentalconnections, and by grant K 125105 of the National Research, Development and
Innovation Fund in Hungary. 

\end{document}